\documentclass[conference]{IEEEtran}
\usepackage{hyperref}
\IEEEoverridecommandlockouts
\usepackage{lipsum}
\usepackage{cite}
\usepackage{amsmath,amssymb,amsfonts}
\usepackage{algorithmic}
\usepackage{graphicx}
\usepackage{textcomp}
\usepackage{xcolor}
\usepackage{booktabs}
\newcommand{\y}{\checkmark}
\def\BibTeX{{\rm B\kern-.05em{\sc i\kern-.025em b}\kern-.08em
    T\kern-.1667em\lower.7ex\hbox{E}\kern-.125emX}}
\begin{document}

\title{HRSim: An agent-based simulation platform for high-capacity ride-sharing services\\

\thanks{This work is jointly supported by Hong Kong PhD Fellowship Scheme (HKPFS) and HKU Research Postgraduate Student Innovation Award.}
}

\author{\IEEEauthorblockN{Wang Chen}
\IEEEauthorblockA{\textit{Dept. of Civil Engineering} \\
\textit{The University of Hong Kong}\\
Hong Kong, China \\
wchen22@connect.hku.hk}
\and
\IEEEauthorblockN{Hongzheng Shi}
\IEEEauthorblockA{\textit{Dept. of Civil Engineering} \\
\textit{The University of Hong Kong}\\
Hong Kong, China \\
shihz@connect.hku.hk}
\and
\IEEEauthorblockN{Jintao Ke}
\IEEEauthorblockA{\textit{Dept. of Civil Engineering} \\
\textit{The University of Hong Kong}\\
Hong Kong, China \\
kejintao@hku.hk}
}

\maketitle

\begin{abstract}
The rapid growth of ride-sharing services presents a promising solution to urban transportation challenges, such as congestion and carbon emissions. However, developing efficient operational strategies—such as pricing, matching, and fleet management—requires robust simulation tools that can replicate real-world dynamics at scale. Existing platforms often lack the capacity, flexibility, or open-source accessibility needed to support large-scale, high-capacity ride-sharing services. To address these gaps, we introduce HRSim, an open-source, agent-based \textit{H}igh-capacity \textit{R}ide-sharing \textit{Sim}ulator. HRSim integrates real-world road networks and demand data to simulate dynamic ride-sharing operations, including pricing, routing, matching, and repositioning. Its module design supports both ride-sharing and solo-hailing service modes. Also, it includes a visualization module for real-time performance analysis. In addition, HRSim incorporates integer linear programming and heuristic algorithms, which can achieve large-scale simulations of high-capacity ride-sharing services. Applications demonstrate HRSim’s utility in various perspectives, including quantifying carbon emissions, scaling ride-sharing performance, evaluating new strategies, etc. By bridging the gap between theoretical research and practical implementation, HRSim serves as a versatile testbed for policymakers and transportation network companies to optimize ride-sharing systems for efficiency and sustainability.

\end{abstract}


\section{Introduction}
In recent years, the rapid pace of urbanization has led to increasingly prominent road traffic issues, including pollution, energy consumption, and congestion. Ride-sharing services enable multiple passengers with similar itineraries to share a trip and have demonstrated positive social impacts in addressing these challenges \cite{Alonso-Mora2017}. At the same time, many transportation network companies (TNCs) such as Uber, Didi, and Lyft have launched ride-sharing services, and we have witnessed a steady growth of ride-sharing services' market share \cite{Kucharski2020}. For example, Uber has expanded ride-sharing services to 11 new markets in 2023 Quarter 4, and the total revenue of shared trips crossed \$1 billion in annual bookings\cite{Uber2023Q4}.

The rapid development of ride-sharing services has raised a series of operational strategy challenges, including pricing, driver-passenger matching, routing, and empty vehicle repositioning. Recently, many advanced algorithms, such as reinforcement learning (RL), have been proposed to address these challenges \cite{Feng2022, Lin2018, Feng2024}. However, the cost of testing these algorithms in real-world ride-sharing operations is prohibitive. Therefore, the development and validation of such algorithms critically depend on a simulation platform that closely replicates the real ride-sharing market \cite{Feng2024}.

The design of ride-sharing simulation platforms presents several challenges. First, the scale of the ride-sharing market necessitates high-capacity simulation platforms capable of accurately capturing market dynamics to support the development of efficient operational strategies. Insufficient capacity can result in inefficient simulations or even short-sighted strategies. Second, the market encompasses a wide range of behaviors, such as typical ride-sharing and solo-hailing. Platforms designed for only one specific operational task are often difficult to extend to other types of operations, thereby limiting their applicability in complex real-world scenarios. Finally, a few simulation platforms are not open source, which restricts the transparency and persuasiveness of the reported effectiveness of many strategies.

\begingroup
\setlength{\tabcolsep}{3pt} % Default value: 6pt
\renewcommand{\arraystretch}{1} % Default value: 1
\begin{table*}[!ht]
    \caption{A comparison between the proposed and existing ride-sharing simulators.} \label{tab: sim_review}
    \begin{center}
    \begin{tabular}{c c c c c c c c c c c c}
    \toprule
    Platform & Open-source & Large-scale & Agent-based & Infrastructure & Pricing & Routing & Matching & Repositioning & Visualization & Solo-hailing & Ride-sharing \\
    \hline
                % & (1) & (2) & (3) & (4) & (5) & (6) & (7) & (8) & (9) & (0) & (1) \\
    UXsim~\cite{uxsim}       & \y  & \y  &     & \y  &     & \y  &     &     & \y  & \y  & \y  \\
    Linares~\cite{b19} &     & \y  &     & \y  &     & \y  &     &     &     &     & \y  \\
    Inturri~\cite{b20} &     & \y  & \y  & \y  &     & \y  & \y  &     &     &     & \y  \\
    D$^2$ABMS~\cite{b21}   &     & \y  & \y  & \y  &     & \y  & \y  & \y  & \y  &     &     \\
    Ding~\cite{b22}    &     &     &     &     &     & \y  & \y  & \y  &     & \y  &     \\
    Hu~\cite{b23}      &     & \y  &     & \y  &     & \y  &     &     & \y  &     &     \\
    FleetPy~\cite{b24}     & \y  & \y  & \y  & \y  & \y  & \y  & \y  & \y  &     & \y  &     \\
    Feng~\cite{Feng2024}    & \y  & \y  & \y  & \y  & \y  & \y  & \y  & \y  & \y  & \y  &     \\
    HRSim & \y & \y & \y & \y & \y & \y & \y & \y & \y & \y & \y \\
    \bottomrule
    \end{tabular}
    \end{center}
\end{table*}

In this paper, we propose HRSim, an agent-based platform for high-capacity ride-sharing simulations. Table~\ref{tab: sim_review} presents a comparison of the functionality of our simulator with that of several existing simulators. Our simulation platform leverages real-world road networks and demand data as its foundation, and jointly simulates pricing, routing, matching, and repositioning processes in ride-sharing services using integer linear programming and heuristic algorithms. The agent-based architecture enables the simulation of multiple behaviors, such as ride-sharing and solo-hailing, and can be easily extended to accommodate additional service modes. Furthermore, we provide an intuitive visualization module that presents simulation results, including passenger statuses, vehicle movements, and a range of detailed statistical indicators. Experimental results demonstrate that our simulator can efficiently perform simulations at a city scale. These features make our simulator a valuable environment for testing high-capacity ride-sharing service strategies and developing advanced algorithms.

\section{Architecture of HRSim}

Fig. \ref{fig: Architecture} depicts the architecture of HRSim, including six modules: infrastructure, pricing, routing, matching, repositioning, and visualization. Specifically, the infrastructure contains road networks, demand, and supply, facilitating ride-sharing simulations in various scenarios. The pricing module determines which passengers will participate in ride-sharing based on their price and detour elasticity calibrated by a questionnaire survey. The routing module calculates the shortest routes using Dijkstra's algorithm \cite{dijksta1959note} and addresses the vehicle routing problem (VRP) by estimating the feasible routes using a heuristic algorithm. The matching module assigns passengers to vehicles by modeling the assignment problem as an integer linear program (ILP). The repositioning module incorporates multiple strategies to dispatch idle vehicles to balance supply and demand. For example, the platform can dispatch idle vehicles to the positions of waiting passengers with the objective of minimizing the total travel distance. Finally, the visualization module records all simulation results and visualizes the movements of all vehicles, based on which the users can check the effectiveness and correctness of their algorithms and strategies. The details of each module are explicitly explained in the next subsections. The source code of HRSim can be accessed at \url{https://github.com/HKU-Smart-Mobility-Lab/Ride-sharing-Simulator}.

\begin{figure*}[!ht]
    \centering
    \includegraphics[width=1.0\linewidth]{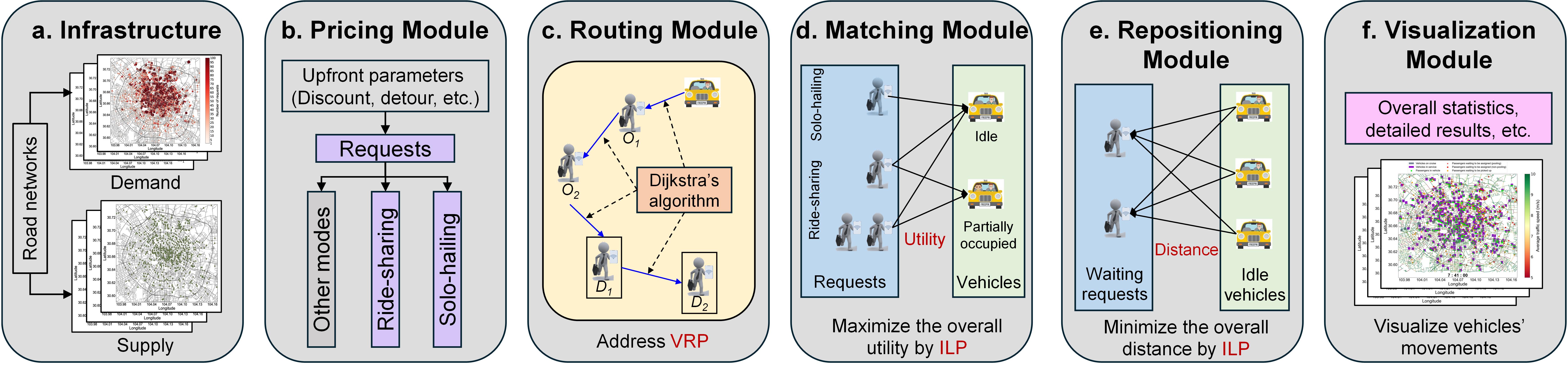}
    \caption{Architecture of HRSim including (a) Infrastructure, (b) pricing module, (c) routing module, (d) matching module, (e)repositioning module, and (f) visualization module.}
    \label{fig: Architecture}
\end{figure*}

\subsection{Infrastructure}

The infrastructure of the simulation platform primarily includes road networks and demand and supply data, as shown in Fig. \ref{fig: Architecture}a. Specifically, road networks can be obtained from Open Street Map \cite{OSM} according to the given area, e.g., longitudinal and latitudinal coordinates. We only extract driver roads since HRSim is implemented to simulate ride-sharing services. Demand data can be accessed from transportation network companies (TNCs) or open-source trip datasets, such as NYC trip data \cite{NYCTrip} and Chicago trip data \cite{ChicagoTrip}. These datasets can generally represent the demand in a city of interest, and thus, they can be used to simulate ride-sharing services. In addition, HRSim can generate trip requests according to a given pattern, such as a uniform distribution or a few hot zones with significantly higher demand, dramatically enhancing the platform's applicability. As for supply data, vehicles are initialized according to the distribution of demand. Vehicles are assumed to be fully compliant and travel according to the routes planned by the platform during simulations. This is reasonable, especially in the era of autonomous driving, as the platform can make decisions based on global information, which can potentially increase the revenue of vehicles if they comply with the platform's management.

\subsection{Pricing module}

The platform incorporates a pricing module to determine which requests are willing to participate in ride-sharing services. Specifically, as shown in Fig. \ref{fig: Architecture}b, the platform charges upfront prices and offers a maximum detour to passengers according to their requests, and passengers can determine whether to accept the service or not. If not, they can opt for solo-hailing services or other transportation modes. It is worth noting that HRSim can simulate solo-hailing and ride-sharing services simultaneously. One can also focus on purely ride-sharing services by excluding solo-hailing services.

We conduct a survey to investigate passengers' price and detour elasticity. In particular, we primally focus on the discount and induced detour of ride-sharing. Fig. \ref{fig: elasticity} depicts the results calibrated with around 400 effective questionnaires. Note that the induced detour, as well as discounts, are relative values compared to solo-hailing. This survey reveals that, when the induced detour ratio is relatively low (e.g., 10\%), many passengers will accept ride-sharing services even if the discount is low. However, passengers' willingness to participate in ride-sharing significantly decreases when the induced detour increases. These results provide a valuable reference for us to design pricing strategies for ride-sharing.

\begin{figure}[!ht]
    \centering
    \includegraphics[width=1.0\linewidth]{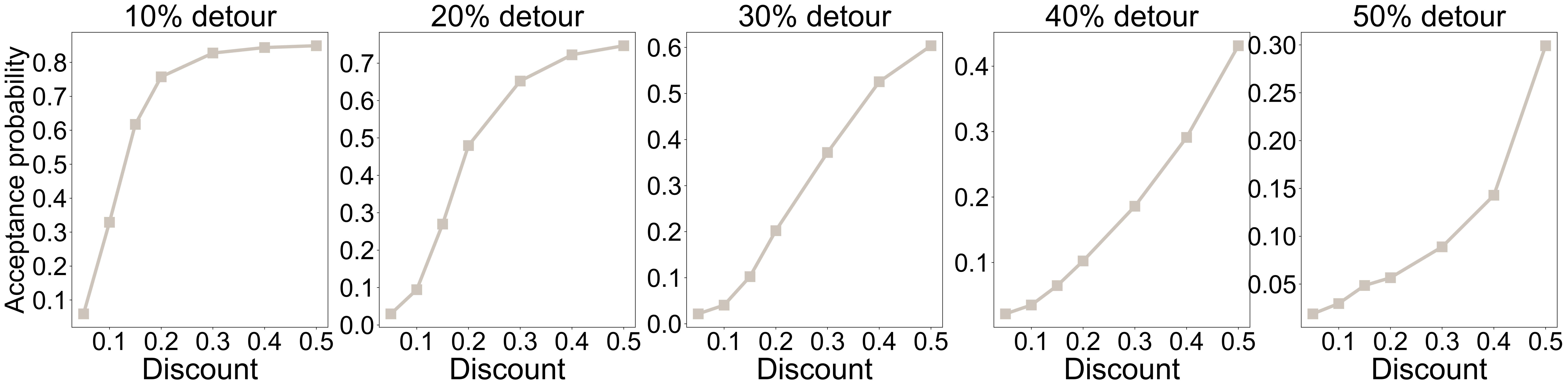}
    \caption{Passengers' Price and detour elasticity calibrated with survey data.}
    \label{fig: elasticity}
\end{figure}

\subsection{Routing module}\label{sec: routing}

To achieve real-time simulations, we decouple routing and matching processes. Specifically, we first connect passengers with vehicles nearby. Then, for all candidate passengers of each vehicle, we determine which passengers can be pooled by checking the feasibility of the shortest route. If the shortest route to pick up and deliver multiple candidate passengers is feasible, i.e., satisfying the maximum detour requirement for all passengers, then these passengers can be pooled as a single trip and matched with the corresponding vehicle. As a result, the platform does not need to consider the feasibility of assigning multiple passengers to one vehicle, dramatically accelerating the assignment process.

One may note that it is time-consuming to calculate the shortest path to accommodate multiple passengers, as the number of possible paths exponentially increases with the number of passengers. To address this issue, we implement a heuristic algorithm, i.e., the nearest neighbor algorithm, to efficiently solve this VRP. Specifically, as shown in Fig. \ref{fig: Architecture}c, HRSim instructs vehicles to iteratively visit their nearest spot to pick up or drop off passengers. This fully greedy algorithm ensures that we can obtain the route on short notice, leading to a real-time simulation even for high-capacity ride-sharing services. Suppose the obtained route cannot meet the requirements, such as inducing a detour exceeding the promised maximum detour ratio or instructing vehicles to visit a passenger's destination before the origin. In that case, the route is infeasible, and these passengers cannot be pooled together. Notably, the effectiveness of this simple method has been verified in \cite{Quant2024Chen}, which can achieve an accuracy of over 95\% compared with the enumeration method.

Vehicles travel from one spot to another along the shortest path instructed by the platform. HRSim adopts Dijkstra's algorithm \cite{dijksta1959note} to calculate the shortest route by modeling the road network as a graph, where roads and intersections correspond to links and nodes, respectively. The weights of the links can be their distances if not considering travel congestion. If taking traffic congestion into account, however, the links' weights should be the travel time estimated based on the traffic flow. In this context, the number of vehicles on each road and their travel velocity should be updated in real time.

\subsection{Matching module}

Once passengers determine whether to participate in ride-sharing, the platform assigns passengers to vehicles nearby. As shown in Fig. \ref{fig: Architecture}d, if passengers choose ride-sharing services, the platform can match them with either idle or partially occupied vehicles. If they opt for solo-hailing services, however, the platform can only match them with idle vehicles. As stated in Sec. \ref{sec: routing}, the routing and matching processes are decoupled. Hence, which trips and vehicles can be potentially matched are determined in the routing module. Furthermore, the utility of each potential match can be calculated since the routes for vehicles to accommodate their trips have been obtained. For example, the assignment utility can be generally defined as the total revenue minus the corresponding costs, as follows:
\begin{equation}
    u_{ij} = \sum_{r \in j}p_r - c_{ij},
\end{equation}
where $u_{ij}$ denotes the utility for matching vehicle $i$ with trip $j$, $p_r$ the charged price from request $r$, and $c_{ij}$ the cost for vehicle $i$ to accommodate trip $j$. It should be noted that a trip may contain one (either solo-hailing or ride-sharing but cannot be pooled with others) or multiple requests. The utility can also be refined to achieve various desired goals. For instance, the utility can contain passengers' waiting times to prioritize matching passengers who have been waiting for a long time. The matching process can be modeled as an ILP. Specifically, let $I$, $R$, and $J$ denote vehicles, requests, and trips with explicit routes for all contained requests, respectively. In addition, let $x_{ij} \in \{0, 1\}$ $( i \in I, j \in J)$ denote the decision variable that determines which vehicle and trip are matched. Therefore, the ILP can be represented as follows:
\begin{align}
    \max_{\mathbf{x}} & {\sum_{i \in I}\sum_{j \in J}x_{ij}u_{ij}} \\
    \text{s.t.} \quad
                    \label{con: 1}
                    & \sum_{j\in J} x_{ij} \leq 1, & \forall i \in I, \\
                    \label{con: 2}
                    & \sum_{i \in I}\sum_{j \in J; r \in j} x_{ij} \leq 1, & \forall r \in R, \\
                    & x_{ij} \in \{0, 1\}, & \forall i \in I, j \in J.
\end{align}
Constraints \eqref{con: 1} and \eqref{con: 2} ensure that each vehicle can be assigned at most one trip and each request can be matched with at most one vehicle, respectively. This program is solved by commercial software in HRSim, e.g., CPLEX or Gurobi. The solution to the above ILP is the final assignment result, based on which vehicles can head to pick up and deliver their passengers according to the planned routes.

\subsection{Repositioning module}

After each matching step, passengers who are not assigned to any vehicles are assumed to continue waiting for future matching steps. HRSim sets a maximum waiting time (e.g., 10 minutes) for passengers that can also be determined by users. Once passengers have been waiting for a longer time, they will cancel their requests and opt for other modes. Regarding idle vehicles that have not been scheduled with any trips during the current matching step, the platform can reposition these vehicles to rebalance the supply. HRSim incorporates multiple repositioning strategies, including cruising nearby and repositioning to waiting passengers. Specifically, the users can instruct idle vehicles to cruise around in a given area (e.g., $2 \times 2$ km) to simulate the repositioning behavior of junior drivers. Also, idle vehicles can stay where they are in HRSim to eliminate the impact of repositioning behavior when testing algorithms designed for the other modules, e.g., matching algorithms.

The platform can also dispatch idle vehicles to waiting passengers' locations to balance the supply and demand. This is modeled as an optimization problem in HRSim to minimize the total repositioning distance, as shown in Fig. \ref{fig: Architecture}e. Specifically, let $V$ and $W$ denote idle vehicles and waiting passengers, respectively. Let $d_{vw}$ $(v \in V, w \in W)$ denote the distance from the location of vehicle $v$ to where passenger $w$ is. Also, let $y_{vw} \in \{0, 1\}$ denote the decision variable determining whether dispatching vehicle $w$ to the location of passenger $w$. Therefore, the repositioning problem can be modeled as an ILP, as follows:
\begin{align}
    \min_{\mathbf{y}} & \sum_{v \in V}\sum_{w \in W} y_{vw}d_{vw} \\
    \text{s.t.} \quad &
                \label{con: 3}
                \sum_{v \in V}y_{vw} \leq 1, & \forall w \in W, \\
                \label{con: 4}
                & \sum_{w \in W}y_{vw} \leq 1, & \forall v \in V, \\
                & y_{vw} \in \{0, 1\}, & \forall v \in V, w \in W.
\end{align}
Constraints \eqref{con: 3} ensure that the platform can dispatch at most one idle vehicle to each waiting passenger's location, and constraints \eqref{con: 4} ensure that each idle vehicle can be dispatched to at most one location. Notably, if the number of idle vehicles is larger than that of waiting passengers, then a portion will stay where they are to reduce the total travel distance. In addition, the location of a few waiting passengers can be the same in hot areas. Thus, the above optimization model can also dispatch multiple idle vehicles to the same location. Furthermore, this optimization model can be adapted to the users' advanced repositioning algorithms by adjusting the repositioning destinations and replacing the travel distance with the users' concerned variables. For example, one may adopt a reinforcement learning (RL) model to dispatch idle vehicles to areas with a high demand in the future. In this context, the distance can be replaced (or added) with the predicted repositioning utility by the RL model.

\subsection{Visualization module}

Once HRSim finishes a simulation, it will output various results, including overall statistics and detailed simulation results. Specifically, HRSim calculates the average service rate of passengers, the induced detour time and distance, the average number of scheduled passengers of each vehicle, vehicle miles traveled, the average revenue, and so on. These overall metrics can help users measure the performance of ride-sharing services, based on which they can adjust their strategies or algorithms to achieve better results. In particular, as shown in Fig. \ref{fig: Architecture}f, HRSim can record and visualize the states of passengers and the movements of vehicles on real-world road networks, facilitating checking the effectiveness and correctness of the simulation processes and results.

\section{Applications of HRSim}

HRSim can be applied to simulate solo-hailing and ride-sharing (or both) services at a city scale. In this section, we introduce three typical applications of HRSim: (1) quantifying carbon emissions, (2) scaling performance metrics, and (3) analyzing pros and cons of strategies.

\subsection{Carbon emission quantification}

One intuitive benefit resulting from ride-sharing services is reducing carbon emissions since the platform can use a smaller fleet to accommodate a given demand. However, it is difficult to quantify to what extent ride-sharing can contribute to carbon emission reductions since the explicit carbon emission data is inaccessible. In this context, HRSim provides a valuable tool for quantifying the carbon emissions of ride-sharing services through simulations. For example, \cite{Quant2024Chen} compared carbon emissions of solo-hailing and ride-sharing with vehicle capacities of 2, 4, and 6 passengers across different service rates, as shown in Fig. \ref{fig: carbon_emissions}. Given a demand, ride-sharing services with a higher vehicle capacity emit less CO$_2$ on average to achieve one kilometer (km) of passenger requests. In particular, compared with solo-hailing services, ride-sharing with vehicle capacities of 2, 4, and 6 passengers can reduce approximately 30\%, 45\%, and 50\% carbon emissions, respectively. These experimental results provide a valuable reference for policymakers and TNCs to better operate ride-sharing services to achieve an environmentally friendly mobility system.

\begin{figure}[!ht]
    \centering
    \includegraphics[width=0.95\linewidth]{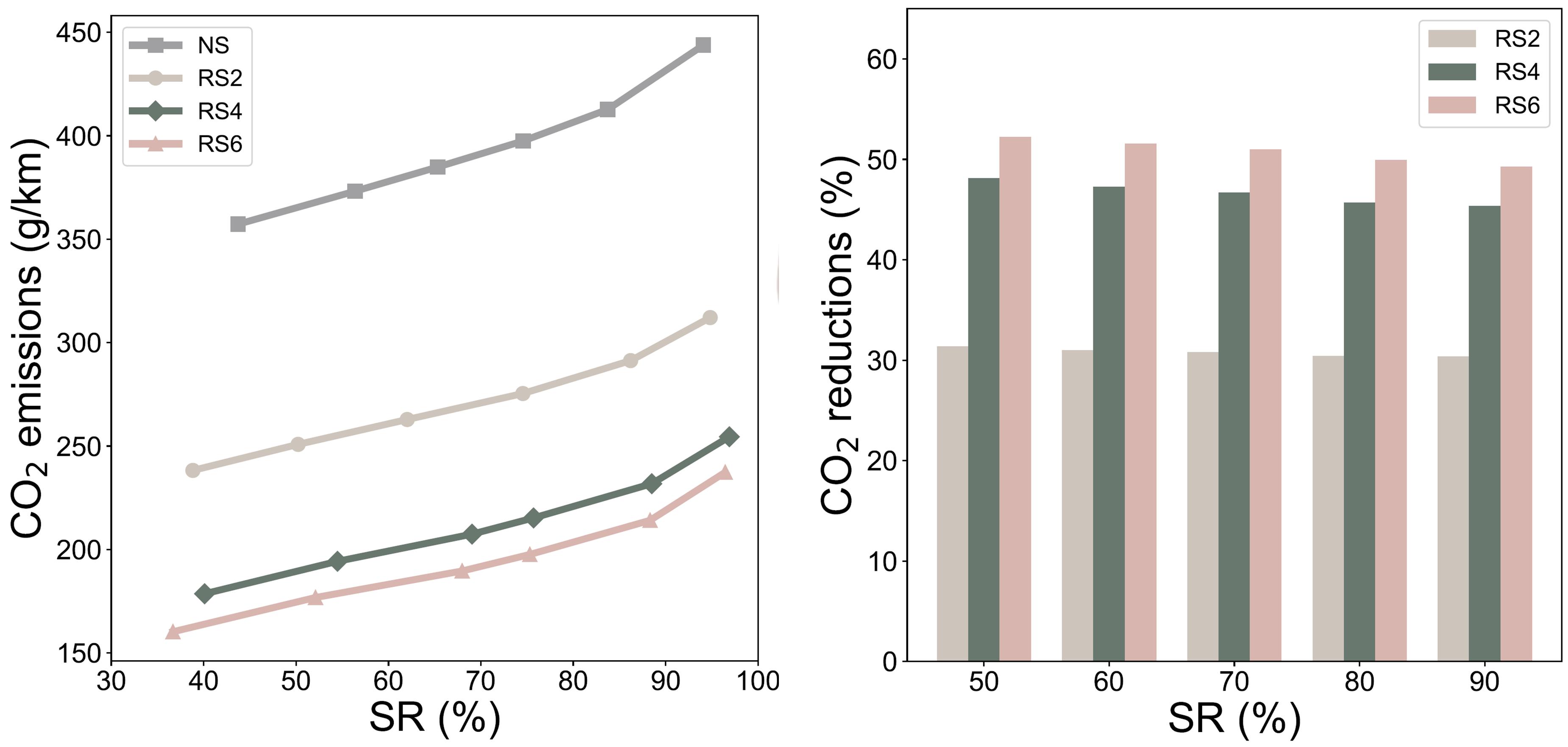}
    \caption{Quantified carbon emissions of non-ride-sharing (NS) and ride-sharing services with vehicle capacities of 2, 4, and 6 passengers (RS2, RS4, and RS6) across different service rates (SR) and the corresponding reduced emissions resulting from ride-sharing.}
    \label{fig: carbon_emissions}
\end{figure}

\subsection{Ride-sharing scaling}

Ride-sharing services, especially high-capacity ride-sharing, are complicated, as the future pooling probability is uncertain. Hence, it is difficult to model a ride-sharing system mathematically. To this end, HRSim can be a useful tool to scale ride-sharing performance based on massive simulations. For example, \cite{Scaling2025Chen} conducted extensive experiments using HRSim, incorporating real-world demand data from ten cities, based on which a few general scaling laws were discovered. In particular, the authors revealed that the key performance metrics, the average service rate of passengers $\Bar{R}$ and the average number of scheduled passengers of each vehicle $\Bar{C}$, can be accurately predicted by only one variable--system load $u$. The system load reflects the demand-to-supply ratio of a ride-sharing system, as follows:
\begin{equation}
    u = \frac{\lambda}{N / \Bar{t}},
\end{equation}
where $\lambda$ denotes the average passenger request arrival rate, $N$ the total fleet size, and $\Bar{t}$ the average service time of passengers, including pickup and delivery time. Furthermore, the scaling laws can be represented as follows:
\begin{equation}
    \Bar{C} = u\Bar{R},
\end{equation}
\begin{equation}
    \label{eq: 2}
    \Bar{R} = \left\{
                \begin{array}{cl}
                    1, & \Bar{C} \leq 1;  \\
                    1-\beta\left( \frac{\Bar{C} -1}{C-1}\right)^\alpha, & \Bar{C} > 1,
                \end{array}
            \right.
\end{equation}
where $C$ denotes the vehicle capacity, and $\alpha$ and $\beta$ are parameters that dependent on $C$: $\alpha = 1.8$ and $\beta = 1.7$ when $C = 2$, $\alpha = 2.7$ and $\beta = 3.5$ when $C = 2$, and $\alpha = 2.4$ and $\beta = 3.6$ when $C = 6$. These scaling laws can accurately reproduce the experimental results across ten cities, indicating that they could be used to predict the performance of ride-sharing services in other cities \cite{Scaling2025Chen}. For instance, when the system load is high (e.g., $u = 3$), high-capacity ride-sharing services with a vehicle capacity of 4 passengers can accommodate 72\% of the demand on average. Low-capacity ride-sharing services with a vehicle capacity of 2 passengers, however, can only accommodate 50\% of the demand according to the scaling laws. These results are crucial for TNCs and urban managers to design ride-sharing strategies, achieving a more efficient mobility system.

\subsection{Strategy analysis}

HRSim can also be used to analyze the effectiveness of the users' proposed strategies or algorithms because (1) HRSim has implemented the entire simulation process and (2) HRSim is an agent-based platform, which ensures that the users can develop their functions based on HRSim without needing to change its architecture. On one hand, users can use HRSim to efficiently validate their pricing, matching, repositioning, or other algorithms, ensuring they can refine and upgrade their algorithms quickly. On the other hand, the users can analyze the potential advantages or disadvantages of incorporating a new policy or strategy based on the simulation results of HRSim. For example, based on extensive simulations conducted on HRSim, \cite{Devel2024Chen} analyzed the social welfare and potential revenue loss of incorporating ride-sharing services compared with pure solo-hailing services. Specifically, the authors used HRSim to simulate pure solo-hailing services and mixed services (including solo-hailing and ride-sharing) involving different discounts and detours. Although the experimental results in different scenarios vary when considering passengers' price and detour elasticity, the authors discovered that, without carefully designing pricing strategies, incorporating ride-sharing services can bring a lot of social benefits but may induce revenue loss. For instance, as shown in Fig. \ref{fig: rs_impact}, compared with pure solo-hailing, incorporating ride-sharing with a 20\% discount and 30\% detour guarantee can increase the service rate by 15.7\%, reduce carbon emissions by 7.1\%, increase the vehicle occupancy rate by 16.7\%, and decrease passengers' waiting time by 2.6\%, respectively. However, incorporating ride-sharing can also lead to a 2.3\% revenue loss if no advanced pricing algorithms are used. These results can guide policymakers to design more efficient strategies for ride-sharing (e.g., providing subsidies), fully leveraging its social benefits to enhance a green and efficient mobility system.

\begin{figure}[!ht]
    \centering
    \includegraphics[width=0.98\linewidth]{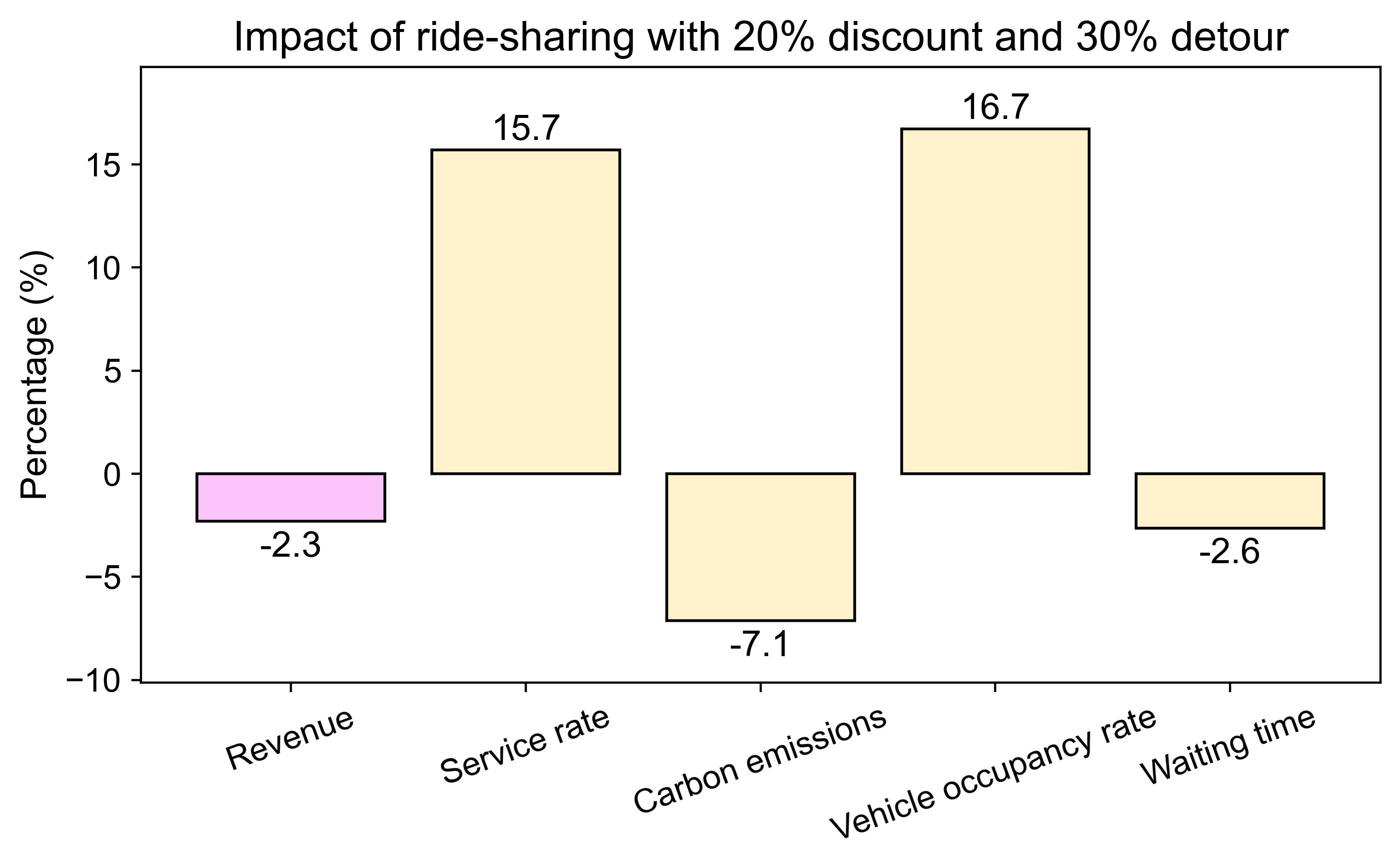}
    \caption{Implications of incorporating ride-sharing services.}
    \label{fig: rs_impact}
\end{figure}

\subsection{Discussion}

HRSim addresses critical research gaps in ride-sharing simulations by providing a scalable, flexible, and open-source platform. Besides the abovementioned applications, HRSim also enables comprehensive testing of advanced algorithms such as RL-based approaches. In addition, its agent-based architecture allows for rapid development of other functions, making it adaptable to complex real-world scenarios. In the future, HRSim could incorporate multi-agent RL or autonomous driving models to further enhance its applicability.

\section{Conclusion}

This paper presents HRSim, a high-capacity, agent-based simulation platform designed to overcome the limitations of existing ride-sharing simulators in scalability, flexibility, and accessibility. By leveraging real-world data and modular processes, HRSim enables the development and validation of advanced operational strategies, from pricing algorithms to fleet management. Its open-source framework ensures transparency and fosters collaboration among researchers and practitioners. HRSim bridges the gap between theoretical research and real-world implementation, offering a powerful tool to optimize ride-sharing systems for efficiency, sustainability, and equity.

\section*{Acknowledgment}

Wang Chen thanks for the support by Hong Kong PhD Fellowship Scheme (HKPFS) and HKU Research Postgraduate Student Innovation Award.

%\section*{References}

\end{document}